\documentclass[sigconf,nonacm]{acmart}

\usepackage{multirow} 
\usepackage{subfigure}
\usepackage{natbib} 
\usepackage{amsmath}
\usepackage{bm}
\usepackage{adjustbox}

\AtBeginDocument{%
  }

\setcopyright{acmcopyright}
\copyrightyear{2027}
\acmYear{2027}
\acmDOI{XXXXXXX.XXXXXXX}

\acmConference[WSDM '27]{the ACM Web Searching and Data Mining 2027}{Feb 15-- Feb 19,
  2027}{HongKong}
\acmPrice{15.00}
\acmISBN{978-1-4503-XXXX-X/XX/XX}

\begin{document}

\title{When Does Supervised Fine-Tuning Reduce Instruction Sensitivity?}


 \author{Jaekeol Choi}
 \email{jaekeol.choi@hufs.ac.kr}
 \affiliation{%
   \institution{Hankuk University of Foreign Studies}
   \city{Seoul}
   \country{South Korea}}





\begin{abstract}
Large language models can exhibit substantial performance variation across alternative formulations of the same task instruction, yet it remains unclear how conventional task-specific supervised fine-tuning (SFT) changes this instruction sensitivity.
We study this question by evaluating fixed model checkpoints under multiple paraphrased instructions and defining instruction sensitivity as the standard deviation of task performance across them.
We conduct a controlled scale analysis with Qwen3 models at 1.7B, 4B, and 8B on MS MARCO, together with targeted cross-family checks using Mistral-7B and Gemma-2-9B.
Before SFT, instruction sensitivity decreases sharply with Qwen3 model scale.
At 1.7B and 4B, SFT consistently reduces sensitivity across training instructions, with reductions of approximately 54--71\%.
At 8B, individual sensitivity changes are not statistically distinguishable from zero, but paired contrasts between training instructions are statistically reliable under query-level bootstrap analysis and have consistent directions across all three random seeds.
Gemma-2-9B shows the same directional training-instruction contrast as Qwen3-8B, whereas Mistral-7B does not, suggesting that the strength of this effect also varies across models.
Experiments on ESCI-English further show that free-generation and likelihood-based forced-choice evaluation can yield qualitatively different robustness conclusions even when valid-label generation is nearly perfect and average task performance is similar.
Overall, SFT does not uniformly reduce instruction sensitivity: its robustness effect depends on the adaptation setting, while measured sensitivity can additionally depend on the prediction and scoring protocol.
\end{abstract}




\maketitle
\section{Introduction}
\label{sec:introduction}

Large language models can exhibit substantial performance variation across alternative formulations of the same task instruction.
Even when two prompts are intended to express the same task, changes in wording, formatting, or specificity can alter model outputs and downstream performance~\cite{webson2022prompt,sclar2024quantifying,sun2024evaluating,mizrahi2024state}.
This issue is particularly relevant to search applications that use LLMs for relevance judgment or reranking, where instruction changes can alter relevance scores and consequently the ranking of retrieved items.
Performance measured with a single instruction may therefore provide an incomplete view of both model effectiveness and deployment reliability.
Recent studies suggest that larger models and supervised or instruction-tuned models tend to be less prompt-sensitive on average~\cite{zhuo2024prosa,qin2026evaluating}, but how instruction sensitivity changes during conventional task-specific supervised fine-tuning (SFT) remains less clear.

Existing approaches to instruction robustness largely focus on explicitly introducing diversity during training.
Large-scale instruction-tuning methods expose models to many tasks and prompt formulations~\cite{wei2022finetuned,sanh2022multitask,mishra2022cross,wang2022supernatural}, while more targeted approaches use paraphrased instructions, contrastive objectives, prompt augmentation, or prompt-agnostic training~\cite{yan2024contrastive,zhao2024selfguide,wei2025paft}.
A more basic setting, however, is common in task-specific adaptation: the model is fine-tuned with a single instruction that remains fixed throughout training.
Whether such ordinary fixed-instruction SFT naturally increases robustness to unseen paraphrases, and whether this effect is consistent across models and scales, is not well understood.

The particular instruction used during fine-tuning may also matter.
Prompt-optimization studies have shown that alternative task formulations can produce substantially different downstream performance~\cite{zhou2023ape,pryzant2023automatic,yang2024optimizers,choi2025efficient}.
More recent work indicates that prompt choice and parameter adaptation can interact, and that fine-tuning may become specialized to particular prompt formulations~\cite{soylu2024finetuning,aissi2025prompt,shi2026training}.
These findings raise a further question: even when different training instructions describe the same task, do they induce the same change in instruction robustness after SFT?

We address these questions through a controlled pre--post analysis of task-specific SFT on passage and product relevance tasks.
For a fixed model checkpoint, we evaluate the same task using a common set of paraphrased instructions designed to preserve the task semantics and define \emph{Instruction Sensitivity} as the standard deviation of task performance across them.
We measure how this sensitivity changes after SFT and compare the changes across alternative fixed training instructions.
Our primary scale-controlled analysis uses Qwen3 models at 1.7B, 4B, and 8B parameters, with targeted cross-family checks using Mistral-7B and Gemma-2-9B.

Experiments on MS MARCO reveal a clear scale-dependent pattern within the Qwen3 family.
Before SFT, instruction sensitivity decreases sharply from 1.7B to 8B.
At 1.7B and 4B, SFT reliably reduces sensitivity to unseen instruction paraphrases under both tested training instructions.
At 8B, however, individual sensitivity changes are not statistically distinguishable from zero, while the paired contrast between training instructions is statistically reliable and consistent across all three random seeds.
A third training instruction produces a similar contrast relative to \(T_A\), showing that the difference is not specific to \(T_B\).
Cross-family checks further qualify this pattern: Gemma-2-9B exhibits the same directional training-instruction contrast as Qwen3-8B, although its confidence interval includes zero, whereas Mistral-7B does not show the same pattern.
Thus, model scale alone is insufficient to explain how SFT changes instruction robustness.

We further investigate whether measured instruction sensitivity depends on the prediction and scoring protocol.
On ESCI-English, free-generation evaluation suggests increased sensitivity after SFT for Qwen3-8B, whereas likelihood-based forced-choice evaluation does not reproduce this pattern.
Average task accuracy remains similar under the two protocols, and valid-label generation exceeds 99.8\% across all conditions.
The discrepancy therefore cannot be explained simply by failures to produce the required output format.
Instead, different prediction procedures can yield qualitatively different conclusions about instruction robustness even when task effectiveness and label compliance are similar, complementing recent concerns about evaluation-induced prompt-sensitivity artifacts~\cite{hua2025flaw}.

Our main contributions are as follows:
\begin{itemize}
    \item We provide a controlled pre--post analysis of instruction sensitivity under conventional fixed-instruction SFT for search relevance tasks.
    \item We show that the robustness effect of SFT varies across model scale and training instruction: Qwen3 exhibits reliable sensitivity reductions at 1.7B and 4B but training-instruction dependence at 8B, while cross-family checks show that the strength and consistency of this effect also vary across models.
    \item We show that measured instruction sensitivity can depend materially on the prediction and scoring protocol, even when average task performance and valid-label generation remain similar.
\end{itemize}

Code, instruction sets, and analysis scripts are available in an anonymous repository.\footnote{\url{https://anonymous.4open.science/r/instruction-sensitivity-sft-61A0}}
\section{Related Work}
\label{sec:related_work}

We review prior work on prompt sensitivity, instruction tuning for robustness, and training-prompt effects in fine-tuning, focusing on how they relate to our controlled pre--post analysis of task-specific SFT.

\subsection{Prompt Sensitivity and Robustness}
\label{sec:rw_prompt_sensitivity}

Language models can exhibit substantial performance variation under prompt changes even when the intended task remains unchanged.
Prior work has documented sensitivity to superficial prompt properties, adversarial perturbations, formatting choices, and alternative task descriptions~\cite{webson2022prompt,zhu2024promptrobust,sclar2024quantifying}.
Such findings raise a practical concern: conclusions drawn from a single prompt may reflect the particular formulation used for evaluation rather than stable task capability.

A particularly relevant setting is variation among natural-language instructions intended to express the same task.
Gu et al.~\cite{gu2023robustness} study instruction paraphrasing and specificity, Sun et al.~\cite{sun2024evaluating} evaluate zero-shot robustness across semantically equivalent instructions, and Mizrahi et al.~\cite{mizrahi2024state} show that prompt choice can affect both absolute performance and model rankings.
Prompt effects have also been observed directly in information retrieval: Sun et al.~\cite{sun2025investigation} show that prompt variations substantially affect zero-shot LLM ranking effectiveness, while Arabzadeh and Clarke~\cite{arabzadeh2025human} analyze prompt sensitivity in LLM-based relevance judgments.
These findings motivate evaluation under multiple instruction formulations.
We therefore focus specifically on \emph{instruction sensitivity}: variation in task performance across alternative natural-language formulations while keeping the underlying task and output space fixed.

Several studies have proposed explicit measures and broader analyses of prompt sensitivity.
POSIX quantifies variability across prompt formulations~\cite{chatterjee2024posix}, while ProSA reports that larger models are generally more stable~\cite{zhuo2024prosa}.
Other work examines sensitivity from calibration and generalization perspectives~\cite{cox2025mapping,liu2026understanding}.
More recently, Qin et al.~\cite{qin2026evaluating} identify model scaling and supervised fine-tuning as factors associated with lower prompt sensitivity across open-source LLMs.

However, lower sensitivity in a fine-tuned model does not by itself reveal how fine-tuning changed robustness.
Most existing analyses compare different pretrained or instruction-tuned checkpoints, model families, or scales rather than measuring the within-model change induced by a particular task-specific adaptation process.
It therefore remains unclear whether SFT consistently reduces instruction sensitivity and whether its effect depends on the instruction used during adaptation.

Prompt-sensitivity estimates can also depend on the evaluation procedure.
Hua et al.~\cite{hua2025flaw} show that apparent sensitivity may partly arise from evaluation artifacts rather than instability in the underlying task decision.
This issue is especially relevant when unrestricted generation and output-based scoring are used, since the prediction and scoring procedure itself can contribute to measured variation.

Our work addresses these gaps through a controlled pre--post SFT analysis.
We evaluate the same model before and after task-specific SFT using a fixed set of paraphrased instructions designed to preserve the task semantics, measure the resulting \emph{change} in instruction sensitivity, and examine how it varies with model scale and the instruction used during fine-tuning.
Targeted comparisons with additional model families further examine whether the observed pattern is specific to the primary scale-controlled model family.

\subsection{Instruction Tuning and Robustness}
\label{sec:rw_instruction_tuning}

Instruction tuning improves the ability of language models to follow natural-language task descriptions and generalize beyond observed tasks.
Natural Instructions~\cite{mishra2022cross}, Super-NaturalInstructions~\cite{wang2022supernatural}, FLAN~\cite{wei2022finetuned}, and T0~\cite{sanh2022multitask} demonstrate that training across diverse task instructions improves cross-task and zero-shot generalization.
These results establish instruction diversity as an important component of model adaptation.

More recent work explicitly targets robustness to instruction variation.
Contrastive Instruction Tuning (CoIN) constructs semantically equivalent instructions and uses contrastive training to improve generalization to unseen formulations~\cite{yan2024contrastive}.
SELF-GUIDE uses self-generated task-specific training data and reports improved instruction following together with reduced sensitivity to prompt formatting~\cite{zhao2024selfguide}.
PAFT addresses prompt-specific overfitting by exposing the model to diverse prompt formulations during fine-tuning, reducing dependence on a particular training prompt~\cite{wei2025paft}.

These studies improve robustness through instruction diversity, specialized objectives, or augmented training data.
Our setting instead uses a single training instruction per SFT run, without an explicit robustness objective or prompt augmentation.
By comparing the same model before and after task-specific SFT across multiple scales, we examine whether standard SFT itself changes robustness to unseen instruction paraphrases and whether this effect depends on the training instruction.

\subsection{Training Prompts and Fine-Tuning}
\label{sec:rw_training_prompts}

A related line of research examines how prompt choice interacts with parameter adaptation.
Prompt-optimization methods such as APE~\cite{zhou2023ape}, ProTeGi~\cite{pryzant2023automatic}, and OPRO~\cite{yang2024optimizers} search for better task instructions while keeping model parameters fixed.
For relevance evaluation, APO-CF similarly optimizes instructions using feedback from model errors~\cite{choi2025efficient}.
These studies show that alternative task formulations can substantially affect downstream performance.

Recent work has begun to study prompt choice together with model fine-tuning.
Soylu et al.~\cite{soylu2024finetuning} show that prompt optimization and weight fine-tuning can provide complementary gains.
Aissi et al.~\cite{aissi2025prompt} show that adaptation to a particular prompt can impair generalization to alternative formulations.
PAFT explicitly varies prompts during fine-tuning to reduce dependence on the training formulation~\cite{wei2025paft}.
More recently, SAPO argues that the fine-tuning prompt can influence the resulting model state and proposes state-adaptive training-prompt selection~\cite{shi2026training}.
Together, these studies suggest that training prompts can shape the outcome of parameter adaptation.

Our work shares this perspective but differs in objective and experimental design.
We do not optimize, augment, or dynamically select the training instruction; each SFT run uses a single instruction and is evaluated on the same paraphrase set with no exact training-instruction overlap.
Rather than asking which prompt yields the highest task performance, we examine how training-instruction choice changes the \emph{robustness effect} of SFT and how this relationship varies with model scale.
\section{Method}
\label{sec:method}

We analyze how task-specific SFT changes performance variation across alternative formulations of the same task instruction and whether this change depends on the instruction used during fine-tuning.

\subsection{Instruction Sensitivity}
\label{sec:instruction_sensitivity}

Let \(\mathcal{I}=\{I_1,\ldots,I_K\}\) denote a set of instruction variants designed to express the same task using different natural-language formulations.
For a fixed model checkpoint \(\theta\), we evaluate the same task instances under each instruction \(I_k\) and obtain the corresponding task performance \(M(\theta,I_k)\).

We define \emph{Instruction Sensitivity} as the sample standard deviation of task performance across the \(K\) instruction variants:

\begin{equation}
S(\theta)
=
\operatorname{Std}_{I_k \in \mathcal{I}}
M(\theta,I_k).
\label{eq:instruction_sensitivity}
\end{equation}

A smaller \(S(\theta)\) indicates that task performance is relatively stable across alternative instruction formulations, whereas a larger value indicates greater dependence on instruction wording.
Because the instruction variants are designed to preserve the same task semantics and output space, the variation captured by \(S(\theta)\) represents performance variation associated with instruction formulation under this controlled instruction set.

Instruction sensitivity is always measured for a fixed model checkpoint while varying only the evaluation instruction.
It is therefore distinct from variation across random seeds or training runs.
For post-SFT models trained with multiple seeds, \(S(\theta)\) is computed separately for each resulting checkpoint, and aggregation across seeds is performed only afterward.

\subsection{SFT-Induced Change in Instruction Sensitivity}
\label{sec:sensitivity_change}

We next quantify how task-specific SFT changes instruction sensitivity.
Let \(S_{\mathrm{pre}}\) denote the sensitivity of the pretrained model and
\(S_{\mathrm{post}}^{T}\) the sensitivity after SFT with training instruction \(T\).
We define the SFT-induced change as

\begin{equation}
\Delta S_T
=
S_{\mathrm{post}}^{T}
-
S_{\mathrm{pre}}.
\label{eq:sensitivity_change}
\end{equation}

The sign of \(\Delta S_T\) directly characterizes the robustness effect of SFT.
A negative value indicates that performance becomes more consistent across instruction variants after fine-tuning, whereas a positive value indicates increased dependence on instruction formulation.
A value near zero indicates little change in instruction sensitivity.

This measure is complementary to conventional task-performance improvement.
SFT may increase average downstream performance while either decreasing or increasing sensitivity across instruction paraphrases.
Accordingly, \(\Delta S_T\) captures how fine-tuning changes robustness to instruction formulation rather than how much it improves the task metric itself.

\subsection{Training-Instruction Dependence}
\label{sec:training_instruction_dependence}

Finally, we examine whether the effect of SFT on instruction sensitivity differs across training instructions.
Given two training instructions \(T_a\) and \(T_b\), we compare their SFT-induced sensitivity changes through

\begin{equation}
D(T_a,T_b)
=
\Delta S_{T_b}
-
\Delta S_{T_a}.
\label{eq:training_instruction_contrast}
\end{equation}

A value of \(D(T_a,T_b)\) close to zero indicates that the two training instructions have similar effects on post-SFT instruction sensitivity.
A positive value indicates that SFT with \(T_b\) results in greater sensitivity than SFT with \(T_a\), relative to the common pretrained baseline, whereas a negative value indicates the opposite.
Because both conditions share the same \(S_{\mathrm{pre}}\), the contrast is equivalently the difference between their post-SFT sensitivities.

This paired contrast is useful because the individual changes \(\Delta S_{T_a}\) and \(\Delta S_{T_b}\) need not be statistically distinguishable from zero for their difference to be reliable.
Accordingly, \(D(T_a,T_b)\) directly tests whether the SFT-induced robustness change differs across training-instruction conditions.
In the main experiments, \(T_A\) and \(T_B\) form the primary comparison, while \(T_C\) is additionally evaluated for Qwen3-8B to further examine the observed training-instruction contrast.
\section{Experimental Setup}
\label{sec:experimental_setup}

We analyze instruction sensitivity across model scales and training instructions using Qwen3 as the primary scale-controlled family, with targeted cross-family checks using Mistral-7B and Gemma-2-9B.

\subsection{Datasets and Tasks}
\label{sec:datasets}

We conduct experiments on two retrieval-oriented tasks: MS MARCO passage ranking and the English subset of the Amazon Shopping Queries Dataset (ESCI).
Table~\ref{tab:datasets} summarizes their roles and evaluation settings.
MS MARCO serves as the primary benchmark for analyzing instruction sensitivity across model scales, training instructions, and model families, while ESCI provides a complementary setting for examining the effect of the evaluation protocol.

\begin{table}[t]
\centering
\caption{Summary of datasets and evaluation settings.}
\label{tab:datasets}
\small
\resizebox{\columnwidth}{!}{
\begin{tabular}{llll}
\hline
Dataset & Task & Evaluation & Primary role \\
\hline
MS MARCO
& Passage ranking
& TREC DL 2019, nDCG@10
& Scale, SFT, and cross-family analysis \\
ESCI-English
& Product relevance
& Accuracy
& Evaluation-protocol analysis \\
\hline
\end{tabular}}
\end{table}

For MS MARCO, we use the passage corpus and training judgments for task-specific fine-tuning and evaluate on the TREC Deep Learning 2019 passage-ranking test set.
For SFT, we sample 5,000 training queries, each contributing one judged relevant passage and one passage sampled from the query's BM25 top-1000 candidates after excluding judged-relevant passages, yielding 10,000 query--passage training instances.
A BM25 retriever provides the top 100 candidate passages for each evaluation query, and all models rerank the same candidate lists.
Each query--passage pair is treated as a binary relevance decision with the labels \textit{Relevant} and \textit{Irrelevant}.

For ESCI, we use the English product-search data and formulate relevance prediction as binary classification.
For SFT, we sample 10,000 training queries, each contributing one positive and one negative product, yielding 20,000 training instances with a balanced 1:1 label ratio.
Products labeled Exact are treated as positive, while Substitute, Complement, and Irrelevant products form the negative class.
Each input consists of a search query and associated product information, and the model predicts either \textit{Relevant} or \textit{Irrelevant}.
ESCI is used primarily to compare free-generation and forced-choice evaluation.

\subsection{Training and Evaluation Instructions}
\label{sec:instructions}

The MS MARCO instruction set is anchored in the relevance-judgment formulation introduced in APO-CF~\cite{choi2025efficient}, which casts passage ranking as a binary decision of whether a candidate passage answers a given query.
For this study, the output space is standardized to the two labels \textit{Relevant} and \textit{Irrelevant}.
Candidate instruction variants were first generated by GPT-5.5 to produce diverse paraphrases of the same relevance-judgment task.
The candidates were then manually screened and, when necessary, revised to preserve the binary query--passage relevance decision and the \textit{Relevant}/\textit{Irrelevant} output space while avoiding additional task requirements or semantic drift.
Importantly, the final training and evaluation instruction sets were fixed before downstream model evaluation and were not selected or modified based on the results reported in this paper.

For MS MARCO, we use three fixed training instructions, denoted \(T_A\), \(T_B\), and \(T_C\).
\(T_A\) and \(T_B\) form the primary comparison and are used for all Qwen3 scales as well as the Mistral-7B and Gemma-2-9B cross-family checks.
\(T_C\) is additionally evaluated for Qwen3-8B to further examine the training-instruction contrast observed at that scale.
Table~\ref{tab:instruction_examples} reports their complete texts together with two representative evaluation instructions.

\begin{table}[t]
\centering
\caption{MS MARCO training instructions and representative evaluation instructions.}
\label{tab:instruction_examples}
\footnotesize
\begin{tabular}{@{}lp{0.91\columnwidth}@{}}
\hline
ID & Instruction \\
\hline

\(T_A\) &
You are a passage relevance judge. Given a search query and a candidate passage, decide whether the passage answers the query.
Respond with exactly one word: ``Relevant'' or ``Irrelevant''. \\[3pt]

\(T_C\) &
You are a passage relevance judge. Given a search query and a candidate passage, decide whether the passage is Relevant or Irrelevant.
Label the passage Relevant only if it directly and clearly answers the query's specific information need.
A passage that only shares the topic, gives background information, or partially touches on the subject without answering it must be labeled Irrelevant.
Respond with exactly one word: ``Relevant'' or ``Irrelevant''. \\[3pt]

\(T_B\) &
You are a passage relevance judge. Given a search query and a candidate passage, decide whether the passage is Relevant or Irrelevant.
Definition: label the passage Relevant ONLY if it directly and substantially answers the query's information need with a clear, dedicated answer---not merely a passage that touches on the same topic.
A passage that is only loosely or tangentially related to the query's subject, or that mentions the topic without actually answering it, must be labeled Irrelevant.
Respond with exactly one word: ``Relevant'' or ``Irrelevant''. \\[3pt]

\(I_1\) &
Does the passage answer the query?
Respond with ``Relevant'' or ``Irrelevant''. \\[3pt]

\(I_8\) &
You are given a search query and a candidate passage.
Your task is to decide whether the passage answers the query.
If the passage answers the query, label it ``Relevant''.
If the passage does not answer the query, label it ``Irrelevant''.
Make a binary decision for each query--passage pair and return exactly one label.
Respond only with ``Relevant'' or ``Irrelevant'', without any additional explanation. \\
\hline
\end{tabular}
\end{table}

The three training instructions differ jointly in length and in the explicitness of the relevance criterion.
\(T_A\) provides the shortest task description, \(T_C\) adds explicit criteria for excluding passages that are merely topical or partially related, and \(T_B\) provides the most detailed relevance definition.
Because these properties co-vary, we treat \(T_A\), \(T_B\), and \(T_C\) as distinct training-instruction conditions rather than attributing their effects to any single linguistic property.

Post-SFT robustness is evaluated using a common set of ten paraphrased instructions, \(I_1,\ldots,I_{10}\).
All ten evaluation instructions are different strings from \(T_A\), \(T_B\), and \(T_C\); thus, there is no exact instruction overlap between training and evaluation.
They were manually screened to preserve the same binary query--passage relevance task and output labels.
The same evaluation instruction set is used across all MS MARCO models, model scales, and SFT conditions, including the Mistral-7B and Gemma-2-9B cross-family checks.

For ESCI-English, we separately construct ten task-equivalent instruction variants, \(E_1,\ldots,E_{10}\).
SFT is conducted with \(E_1\) and \(E_{10}\), and all ten variants are used during evaluation.
Because the ESCI experiments primarily examine the effect of the output-evaluation protocol, the training instructions are also included among the evaluation variants rather than enforcing the no-exact-overlap design used for MS MARCO.

The complete instruction sets and experimental configurations are available in the anonymous repository.

\subsection{Models and Fine-Tuning Setup}
\label{sec:training_setup}

Our primary scale-controlled analysis uses Qwen3 models with 1.7B, 4B, and 8B parameters.
Using models from the same family allows us to examine scale-dependent changes while minimizing architectural differences.
To assess whether the behavior observed at the larger Qwen3 scale extends beyond this family, we additionally evaluate Mistral-7B and Gemma-2-9B as targeted cross-family checks.

All models are initialized from publicly released instruction-tuned checkpoints rather than raw base models.
These checkpoints serve as the pretrained reference before our task-specific SFT.
We adapt each model with LoRA (\(r=8\), \(\alpha=16\)) using AdamW with a learning rate of \(2\times10^{-5}\), a warmup ratio of 0.1, an effective batch size of 8, a maximum sequence length of 512, and three training epochs.
The same fine-tuning configuration is used across model families and training-instruction conditions, without model-specific hyperparameter tuning.
Within each dataset, the same training instances are used across training-instruction conditions and random seeds; the seeds affect LoRA initialization and batch-order shuffling, but not the training subset.
Inputs are formatted using each model's native chat template while keeping the instruction text, task inputs, and output labels unchanged.
Post-SFT results are reported over three random seeds, while each pretrained checkpoint is evaluated once as the corresponding Pre-SFT reference.
Exact checkpoint identifiers, complete training configurations, and implementation details are provided in the anonymous repository.

For MS MARCO, \(T_A\) and \(T_B\) are used for all three Qwen3 scales and for the Mistral-7B and Gemma-2-9B cross-family checks.
\(T_C\) is additionally evaluated for Qwen3-8B.
The ESCI protocol analysis is conducted with Qwen3-8B checkpoints trained using \(E_1\) and \(E_{10}\).

\subsection{Evaluation and Statistical Analysis}
\label{sec:evaluation}

We evaluate both task effectiveness and robustness to instruction variation.
For each checkpoint, task performance is measured under all ten evaluation instructions, and instruction sensitivity is computed as defined in Section~\ref{sec:method}.
For Pre-SFT models, \(S\) is the sample standard deviation across the ten instruction-level scores.
For Post-SFT models, \(S\) is computed separately for each seed and then averaged over three seeds; per-instruction performance is reported as mean$\pm$standard deviation across seeds.

For MS MARCO, ranking scores are computed without free-form generation.
For each query--passage pair, we compute the full conditional sequence log-likelihood of \textit{Relevant} and \textit{Irrelevant}, including all tokens in each label, and use
\(s(q,p)=\log P(\textit{Relevant}\mid q,p)-\log P(\textit{Irrelevant}\mid q,p)\)
as the ranking score.
Candidate passages are ranked by \(s(q,p)\), and effectiveness is measured using nDCG@10.
The same likelihood-based scoring procedure is used for all Qwen3, Mistral-7B, and Gemma-2-9B conditions.

For ESCI-English, we compare free-generation and forced-choice evaluation.
Under free generation, the model uses deterministic greedy decoding with a maximum of six generated tokens.
After stripping surrounding whitespace, the output is mapped to \textit{Relevant} or \textit{Irrelevant} using case-sensitive prefix matching.
Under forced choice, the label with the higher full conditional sequence likelihood is selected directly.
These protocols therefore differ in how task predictions are obtained and scored.
For free generation, we additionally report the valid-label rate to determine whether protocol differences can be explained by output-format failures.

For all MS MARCO comparisons, statistical reliability is assessed using paired query-level bootstrap resampling with 10,000 replicates.
Within each model, the same sampled query indices are used across evaluation instructions, training conditions, and random seeds in each replicate, preserving the paired structure of the comparison.
Within each replicate, \(S\) is recomputed separately for each seed and then averaged across seeds before forming \(\Delta S\) and \(D\).
Reported point estimates of \(\Delta S\) and \(D\) are computed from the original, non-resampled evaluation set, while the bootstrap replicates are used to obtain percentile-based 95\% confidence intervals and, for directional contrasts, the empirical probability \(P(D>0)\).
This analysis quantifies query-level uncertainty while treating the trained seeds as observed runs.
For the larger-model training-instruction contrasts, we additionally examine seed-level consistency and repeat the paired bootstrap while leaving out one seed at a time.
\section{Experimental Results}
\label{sec:results}

We first examine how instruction sensitivity varies with model scale in the primary Qwen3 analysis and then analyze how task-specific SFT changes this sensitivity under different training instructions.
We further assess the reliability and cross-family behavior of the observed training-instruction effect using Mistral-7B and Gemma-2-9B, and finally examine how the evaluation protocol influences measured instruction sensitivity.

\subsection{Performance Across Instruction Paraphrases}
\label{sec:results_instruction_performance}

We first examine the primary Qwen3 results across alternative formulations of the same relevance task.
Table~\ref{tab:instruction_performance} reports nDCG@10 for ten paraphrased evaluation instructions, denoted \(I_1\)--\(I_{10}\), with no exact instruction overlap with the training conditions.
For the SFT models, each entry reports the mean and standard deviation over three random seeds.
The same evaluation instructions are used across all Qwen3 model scales and SFT conditions.

\begin{table*}[t]
\centering
\caption{nDCG@10 across ten paraphrased evaluation instructions on MS MARCO with no exact instruction overlap with the training conditions.
Post-SFT results are reported as mean$\pm$standard deviation over three seeds.
\(S\) denotes instruction sensitivity, computed as the standard deviation of performance across the ten evaluation instructions.}
\label{tab:instruction_performance}
\small
\resizebox{\textwidth}{!}{
\begin{tabular}{lccc|ccc|ccc}
\hline
& \multicolumn{3}{c|}{Qwen3-1.7B}
& \multicolumn{3}{c|}{Qwen3-4B}
& \multicolumn{3}{c}{Qwen3-8B} \\
\cline{2-10}
Eval. inst.
& Pre-SFT & SFT-$T_A$ & SFT-$T_B$
& Pre-SFT & SFT-$T_A$ & SFT-$T_B$
& Pre-SFT & SFT-$T_A$ & SFT-$T_B$ \\
\hline
$I_1$  & 0.2003 & $0.4801{\pm}0.0151$ & $0.4547{\pm}0.0191$
       & 0.6535 & $0.7174{\pm}0.0065$ & $0.7046{\pm}0.0128$
       & 0.7019 & $0.7268{\pm}0.0010$ & $0.6900{\pm}0.0202$ \\

$I_2$  & 0.2687 & $0.4833{\pm}0.0122$ & $0.4775{\pm}0.0034$
       & 0.6382 & $0.7034{\pm}0.0072$ & $0.6851{\pm}0.0136$
       & 0.6933 & $0.7295{\pm}0.0054$ & $0.7117{\pm}0.0086$ \\

$I_3$  & 0.4420 & $0.5281{\pm}0.0070$ & $0.4815{\pm}0.0065$
       & 0.6393 & $0.6991{\pm}0.0012$ & $0.6936{\pm}0.0047$
       & 0.7211 & $0.7407{\pm}0.0024$ & $0.7151{\pm}0.0059$ \\

$I_4$  & 0.3997 & $0.5234{\pm}0.0032$ & $0.5045{\pm}0.0133$
       & 0.6528 & $0.7305{\pm}0.0022$ & $0.7224{\pm}0.0053$
       & 0.6887 & $0.7457{\pm}0.0028$ & $0.7426{\pm}0.0028$ \\

$I_5$  & 0.3855 & $0.5547{\pm}0.0023$ & $0.5474{\pm}0.0048$
       & 0.6351 & $0.7411{\pm}0.0034$ & $0.7279{\pm}0.0070$
       & 0.7174 & $0.7439{\pm}0.0037$ & $0.7459{\pm}0.0007$ \\

$I_6$  & 0.2879 & $0.5158{\pm}0.0032$ & $0.4738{\pm}0.0115$
       & 0.5684 & $0.7211{\pm}0.0023$ & $0.6962{\pm}0.0080$
       & 0.7138 & $0.7511{\pm}0.0020$ & $0.7410{\pm}0.0095$ \\

$I_7$  & 0.4754 & $0.5424{\pm}0.0108$ & $0.5287{\pm}0.0022$
       & 0.6041 & $0.7335{\pm}0.0049$ & $0.7165{\pm}0.0070$
       & 0.7031 & $0.7373{\pm}0.0013$ & $0.7292{\pm}0.0078$ \\

$I_8$  & 0.4154 & $0.5469{\pm}0.0093$ & $0.5221{\pm}0.0121$
       & 0.6639 & $0.7276{\pm}0.0012$ & $0.7089{\pm}0.0060$
       & 0.7190 & $0.7475{\pm}0.0031$ & $0.7450{\pm}0.0040$ \\

$I_9$  & 0.4268 & $0.5214{\pm}0.0046$ & $0.5040{\pm}0.0150$
       & 0.6372 & $0.7283{\pm}0.0038$ & $0.7132{\pm}0.0024$
       & 0.7107 & $0.7338{\pm}0.0034$ & $0.7296{\pm}0.0019$ \\

$I_{10}$ & 0.4429 & $0.5481{\pm}0.0025$ & $0.5207{\pm}0.0146$
         & 0.6907 & $0.7376{\pm}0.0046$ & $0.7281{\pm}0.0046$
         & 0.7274 & $0.7484{\pm}0.0022$ & $0.7434{\pm}0.0044$ \\
\hline
Mean
       & 0.3745 & 0.5244 & 0.5015
       & 0.6383 & 0.7240 & 0.7096
       & 0.7096 & 0.7405 & 0.7293 \\
Sensitivity \(S\)
       & 0.0905 & 0.0265 & 0.0299
       & 0.0332 & 0.0143 & 0.0152
       & 0.0126 & 0.0086 & 0.0189 \\
\hline
\end{tabular}}
\end{table*}

Before SFT, performance varies substantially across instruction paraphrases for the smaller models.
Qwen3-1.7B ranges from 0.2003 to 0.4754 nDCG@10 across the ten instructions, resulting in an instruction sensitivity of \(S=0.0905\).
The variation becomes considerably smaller as model scale increases: \(S\) decreases to \(0.0332\) for Qwen3-4B and to \(0.0126\) for Qwen3-8B.
Thus, within the tested Qwen3 family, larger pretrained models are not only more effective on average but also substantially more stable across alternative formulations of the same task instruction.

SFT improves mean ranking effectiveness at all three model scales.
For Qwen3-1.7B, mean nDCG@10 increases from 0.3745 before SFT to 0.5244 under \(T_A\) and 0.5015 under \(T_B\).
For Qwen3-4B, the corresponding mean increases from 0.6383 to 0.7240 and 0.7096, while Qwen3-8B improves from 0.7096 to 0.7405 and 0.7293.
The per-instruction results also show that the effect of SFT on performance variation is not identical across model scales and training instructions.
We examine these changes in instruction sensitivity directly in the next section.

\subsection{SFT-Induced Changes in Instruction Sensitivity}
\label{sec:results_sft_change}

\begin{figure}[t]
    \centering
    \includegraphics[width=0.8\linewidth]{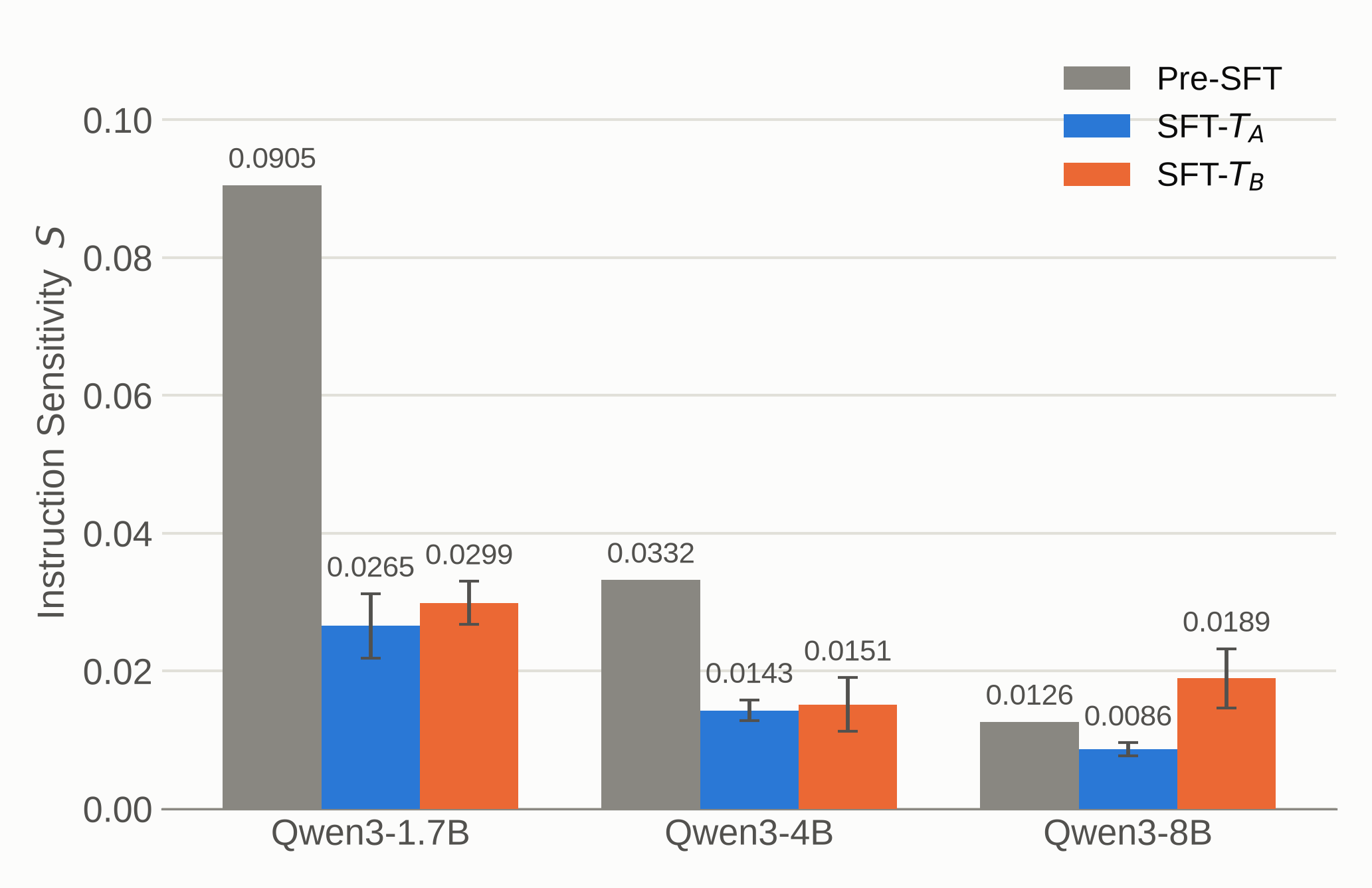}
    \caption{Instruction sensitivity before and after task-specific SFT on MS MARCO for Qwen3.
    Sensitivity \(S\) is the standard deviation of nDCG@10 across ten paraphrased evaluation instructions with no exact overlap with the training instructions.
    Post-SFT values are averaged over three random seeds.}
    \label{fig:sensitivity_scale}
\end{figure}

We next examine how task-specific SFT changes sensitivity to instruction paraphrases.
We consider two training instructions, denoted \(T_A\) and \(T_B\), and evaluate each resulting model using the same ten paraphrased instructions used in Section~\ref{sec:results_instruction_performance}, with no exact instruction overlap between training and evaluation.
Figure~\ref{fig:sensitivity_scale} summarizes instruction sensitivity before and after SFT across the three Qwen3 model scales.

For Qwen3-1.7B, pre-SFT sensitivity is \(0.0905\).
After SFT, it decreases to \(0.0265\) with \(T_A\) and \(0.0299\) with \(T_B\), corresponding to relative reductions of \(70.7\%\) and \(67.0\%\), respectively.
A similar pattern is observed for Qwen3-4B: sensitivity decreases from \(0.0332\) to \(0.0143\) with \(T_A\) and to \(0.0152\) with \(T_B\), representing reductions of \(57.1\%\) and \(54.3\%\).
Thus, at both smaller Qwen3 scales, task-specific SFT substantially compresses performance variation across unseen instruction paraphrases under both training instructions.

The behavior changes at Qwen3-8B.
The pretrained model is already considerably less sensitive, with \(S=0.0126\).
SFT with \(T_A\) further decreases sensitivity to \(0.0086\), a \(31.3\%\) reduction.
In contrast, SFT with \(T_B\) produces a sensitivity of \(0.0189\), corresponding to a \(50.4\%\) increase relative to the pre-SFT value.
Accordingly, the two training instructions that produce similar sensitivity reductions at 1.7B and 4B exhibit qualitatively different point-estimate behavior at 8B.

The scale-wise pattern within Qwen3 is also informative.
For \(T_A\), the relative reduction becomes progressively smaller, from \(70.7\%\) at 1.7B to \(57.1\%\) at 4B and \(31.3\%\) at 8B.
For \(T_B\), the reduction similarly weakens from \(67.0\%\) to \(54.3\%\) before changing direction at 8B.
These results show that the robustness effect of SFT is not uniform across Qwen3 model scales and motivate examining whether the apparent training-instruction dependence at 8B is statistically reliable.

The 8B values in Figure~\ref{fig:sensitivity_scale}, however, are point estimates and do not by themselves establish that either the decrease under \(T_A\) or the increase under \(T_B\) is statistically reliable.
In Section~\ref{sec:results_train_inst}, we therefore test the training-instruction contrast directly using paired query-level bootstrap analysis and further examine whether the observed pattern extends beyond the Qwen3 family.

\subsection{Training-Instruction Dependence and Cross-Model Comparison}
\label{sec:results_train_inst}

The divergent Qwen3-8B pattern raises the question of whether the effect of SFT on instruction sensitivity differs across training instructions.
We assess this using paired query-level bootstrap analysis with 10,000 resamples.
Within each replicate, the same sampled queries are used across evaluation instructions, training conditions, and random seeds.
This procedure quantifies query-level uncertainty while treating the trained runs as observed checkpoints.
Table~\ref{tab:bootstrap_sensitivity} reports the individual SFT-induced changes \(\Delta S\) for the primary Qwen3 analysis.

\begin{table}[t]
\centering
\caption{Point estimates of individual SFT-induced changes in instruction sensitivity for Qwen3 on MS MARCO, with 95\% confidence intervals from 10,000 paired query-level bootstrap resamples.
Negative \(\Delta S\) indicates reduced sensitivity after SFT.}
\label{tab:bootstrap_sensitivity}
\small
\begin{tabular}{llcc}
\hline
Model & Training inst. & $\Delta S$ & 95\% CI \\
\hline
Qwen3-1.7B
& $T_A$ & $-0.0640$ & $[-0.0846,-0.0418]$ \\
& $T_B$ & $-0.0607$ & $[-0.0781,-0.0407]$ \\
\hline
Qwen3-4B
& $T_A$ & $-0.0189$ & $[-0.0339,-0.0072]$ \\
& $T_B$ & $-0.0180$ & $[-0.0322,-0.0051]$ \\
\hline
Qwen3-8B
& $T_A$ & $-0.0039$ & $[-0.0151,+0.0032]$ \\
& $T_B$ & $+0.0063$ & $[-0.0073,+0.0166]$ \\
\hline
\end{tabular}
\end{table}

At 1.7B and 4B, both training instructions yield statistically reliable reductions in instruction sensitivity, with all confidence intervals lying below zero.
At 8B, however, the point estimate is negative for \(T_A\) and positive for \(T_B\), while both confidence intervals include zero.
Thus, neither the decrease under \(T_A\) nor the increase under \(T_B\) is statistically reliable when considered individually.

\begin{figure}[t]
    \centering
    \includegraphics[width=0.8\linewidth]{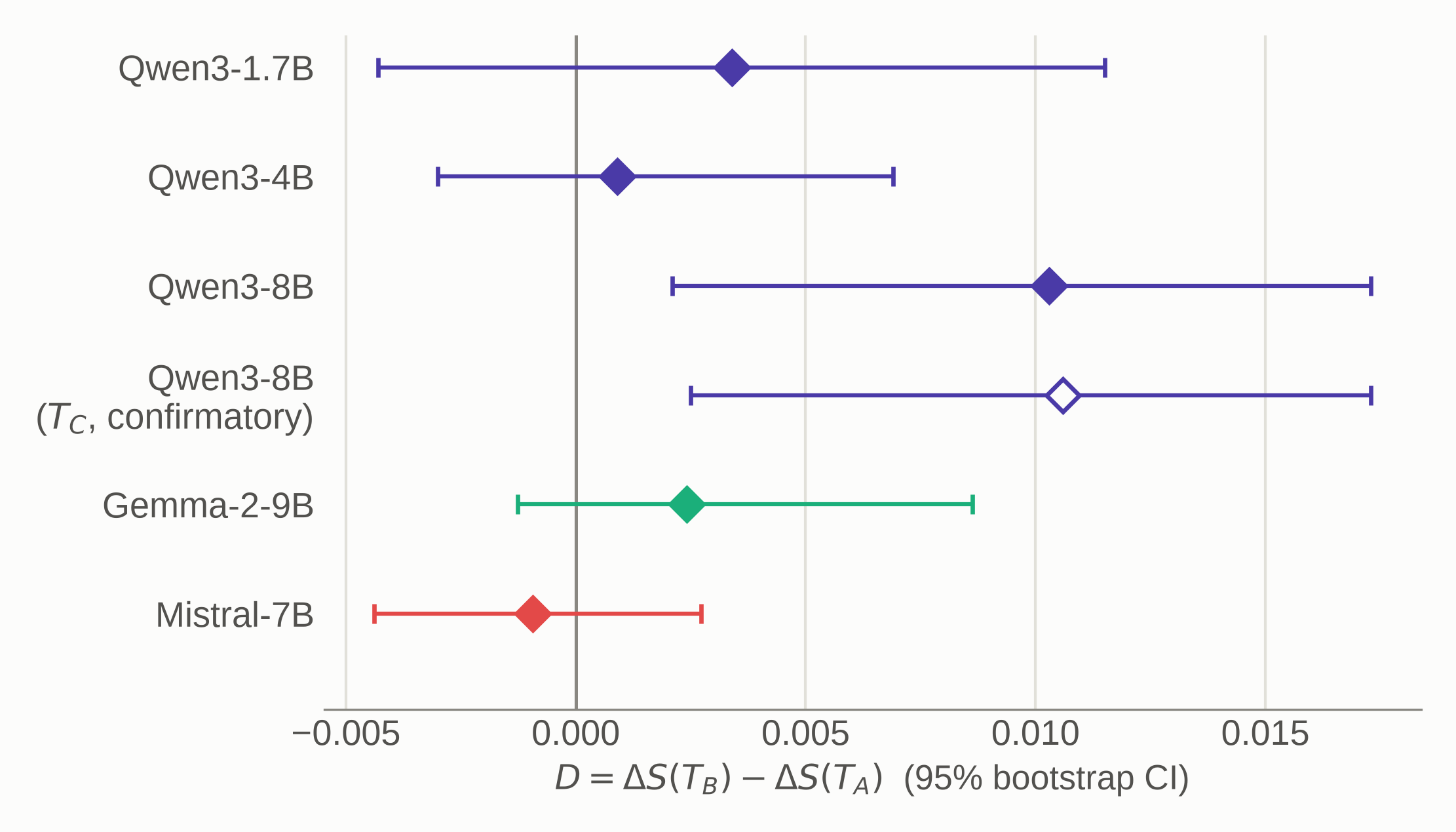}
    \caption{Paired training-instruction contrasts in SFT-induced instruction sensitivity on MS MARCO.
    Points show the observed contrast \(D=\Delta S(T_B)-\Delta S(T_A)\) computed on the original evaluation set, and error bars indicate 95\% paired query-bootstrap confidence intervals.
    Qwen3-8B exhibits a statistically reliable positive contrast, while Gemma-2-9B shows the same directional pattern without excluding zero and Mistral-7B shows no corresponding contrast.
    The open marker denotes the additional \(T_C-T_A\) comparison for Qwen3-8B.}
    \label{fig:training_instruction_contrast}
\end{figure}

We therefore examine the paired training-instruction contrast
\(D=\Delta S(T_B)-\Delta S(T_A)\), which directly tests whether the robustness effect of SFT differs between the two training conditions.
Figure~\ref{fig:training_instruction_contrast} summarizes these contrasts.
For Qwen3-1.7B, \(D=0.0034\) with a 95\% confidence interval of
\([-0.0043,0.0115]\), while Qwen3-4B yields \(D=0.0009\) with
\([-0.0030,0.0069]\).
Neither contrast is statistically distinguishable from zero.
In contrast, Qwen3-8B yields \(D=0.0103\), with
\([0.0021,0.0173]\) and \(P(D>0)=0.996\).
Thus, although the individual 8B changes are uncertain, their paired difference is statistically reliable with respect to query-level variation.

To examine whether the Qwen3-8B contrast relative to \(T_A\) also appears with another training instruction, we additionally evaluate \(T_C\).
The \(T_C-T_A\) contrast is \(D=0.0106\), with a 95\% confidence interval of
\([0.0025,0.0173]\) and \(P(D>0)=0.997\), closely matching the \(T_B-T_A\) result.
Thus, the contrast relative to \(T_A\) is observed with more than one alternative training instruction.
However, with only three training instructions, the analysis does not establish whether \(T_A\) itself is exceptional or which instruction properties drive the difference.

We further assess whether the Qwen3-8B contrasts are driven by a particular random seed.
For \(T_B-T_A\), the seed-level contrasts are \(0.0116\), \(0.0064\), and \(0.0128\); for \(T_C-T_A\), they are \(0.0107\), \(0.0070\), and \(0.0142\).
Both contrasts are therefore positive across all three observed seeds.
Moreover, all six leave-one-seed-out paired-bootstrap analyses retain positive contrasts with 95\% confidence intervals entirely above zero, indicating that the query-level significance is not driven by any single training run.

Finally, we examine whether the Qwen3-8B pattern extends to other models at comparable scales.
For Gemma-2-9B, sensitivity decreases from \(S_{\mathrm{Pre}}=0.0086\) to \(S_{T_A}=0.0059\), whereas \(S_{T_B}=0.0083\) remains close to the Pre-SFT level, yielding \(D=0.0024\).
The contrast is positive in all three seeds and remains positive in every leave-one-seed-out analysis.
However, its paired bootstrap confidence interval, \([-0.0013,0.0086]\), includes zero, so Gemma-2-9B provides the same directional pattern without a statistically reliable contrast.
Mistral-7B does not show the same pattern: relative to \(S_{\mathrm{Pre}}=0.0174\), sensitivity decreases to \(S_{T_A}=0.0165\) and \(S_{T_B}=0.0155\), giving \(D=-0.0009\) with a confidence interval of \([-0.0044,0.0027]\).

Overall, the Qwen3 analysis shows that SFT with either \(T_A\) or \(T_B\) reliably reduces instruction sensitivity at 1.7B and 4B, whereas the effect differs across training instructions at 8B.
The cross-model checks provide a more qualified picture: Gemma-2-9B exhibits the same directional contrast as Qwen3-8B, whereas Mistral-7B does not.
These results suggest that model scale alone is insufficient to determine how SFT affects instruction robustness; the magnitude and consistency of training-instruction dependence can also vary across models.

\subsection{Evaluation Protocol and Apparent Instruction Sensitivity}
\label{sec:results_protocol}

Finally, we examine whether measured instruction sensitivity depends on the evaluation protocol.
This analysis uses ESCI-English with Qwen3-8B and compares two prediction procedures.
Under \emph{free generation}, the model produces an output using deterministic greedy decoding and maps it to a label using case-sensitive prefix matching.
Under \emph{forced choice}, we compute the full conditional log-likelihoods of \textit{Relevant} and \textit{Irrelevant} and select the higher-scoring label.
Because these procedures differ in how predictions are obtained and scored, their differences cannot be attributed solely to output-format failures.

Table~\ref{tab:esci_protocol_results} reports per-instruction accuracy under both protocols.
Post-SFT results are reported as mean$\pm$standard deviation over three random seeds.

\begin{table}[t]
\centering
\caption{Per-instruction accuracy on ESCI-English with Qwen3-8B under free-generation and forced-choice evaluation.
\(E_1\)--\(E_{10}\) denote the ten evaluation instructions.
Post-SFT values are mean$\pm$standard deviation over three seeds, and \(S\) denotes instruction sensitivity.}
\label{tab:esci_protocol_results}
\resizebox{\columnwidth}{!}{
\begin{tabular}{lccc|ccc}
\hline
&
\multicolumn{3}{c|}{Free generation} &
\multicolumn{3}{c}{Forced choice} \\
\cline{2-7}
Eval. inst.
& Pre-SFT & SFT-$E_1$ & SFT-$E_{10}$
& Pre-SFT & SFT-$E_1$ & SFT-$E_{10}$ \\
\hline
$E_1$  & 0.6483 & $0.7028{\pm}0.0085$ & $0.6878{\pm}0.0068$
       & 0.6450 & $0.6961{\pm}0.0026$ & $0.6861{\pm}0.0035$ \\
$E_2$  & 0.6600 & $0.6661{\pm}0.0122$ & $0.6422{\pm}0.0079$
       & 0.6583 & $0.6656{\pm}0.0096$ & $0.6522{\pm}0.0100$ \\
$E_3$  & 0.6383 & $0.6572{\pm}0.0017$ & $0.6628{\pm}0.0025$
       & 0.6333 & $0.6644{\pm}0.0051$ & $0.6617{\pm}0.0017$ \\
$E_4$  & 0.6700 & $0.6728{\pm}0.0053$ & $0.6717{\pm}0.0076$
       & 0.6683 & $0.6717{\pm}0.0093$ & $0.6739{\pm}0.0063$ \\
$E_5$  & 0.6383 & $0.6611{\pm}0.0122$ & $0.6539{\pm}0.0090$
       & 0.6367 & $0.6672{\pm}0.0084$ & $0.6600{\pm}0.0104$ \\
$E_6$  & 0.6633 & $0.6733{\pm}0.0035$ & $0.6789{\pm}0.0017$
       & 0.6617 & $0.6778{\pm}0.0039$ & $0.6739{\pm}0.0048$ \\
$E_7$  & 0.6467 & $0.6594{\pm}0.0108$ & $0.6667{\pm}0.0042$
       & 0.6483 & $0.6622{\pm}0.0155$ & $0.6656{\pm}0.0079$ \\
$E_8$  & 0.6650 & $0.6678{\pm}0.0050$ & $0.6583{\pm}0.0076$
       & 0.6683 & $0.6717{\pm}0.0101$ & $0.6706{\pm}0.0042$ \\
$E_9$  & 0.6667 & $0.6750{\pm}0.0062$ & $0.6844{\pm}0.0060$
       & 0.6633 & $0.6778{\pm}0.0019$ & $0.6828{\pm}0.0058$ \\
$E_{10}$ & 0.6650 & $0.6828{\pm}0.0045$ & $0.6961{\pm}0.0035$
         & 0.6700 & $0.6861{\pm}0.0035$ & $0.6967{\pm}0.0067$ \\
\hline
Mean
& 0.6562 & 0.6718 & 0.6703
& 0.6553 & 0.6741 & 0.6723 \\
Sensitivity \(S\)
& 0.0121 & 0.0141 & 0.0171
& 0.0136 & 0.0119 & 0.0138 \\
\hline
\end{tabular}}
\end{table}

The two protocols yield similar conclusions about overall task performance.
Mean accuracy increases after SFT from approximately \(0.656\) to \(0.672\) and \(0.670\) under free generation, and from \(0.6553\) to \(0.6741\) and \(0.6723\) under forced choice.
Thus, both protocols indicate improved average ESCI performance after SFT.

Their conclusions about instruction sensitivity, however, differ.
Under free generation, sensitivity increases from \(0.0121\) to \(0.0141\) and \(0.0171\).
Under forced choice, it instead changes from \(0.0136\) to \(0.0119\) and \(0.0138\).
The apparent post-SFT increase under free generation is therefore not reproduced under likelihood-based forced choice.

This discrepancy cannot be explained simply by invalid output formatting.
Across Pre-SFT, SFT-\(E_1\), and SFT-\(E_{10}\), the mean valid-label rate is \(0.9983\), with a standard deviation of only \(0.0027\) across evaluation instructions; individual rates range from \(0.9933\) to \(1.0000\).
Thus, Qwen3-8B follows the required label format almost perfectly even when the two protocols yield different sensitivity patterns.

\begin{figure}[t]
    \centering
    \includegraphics[width=0.9\linewidth]{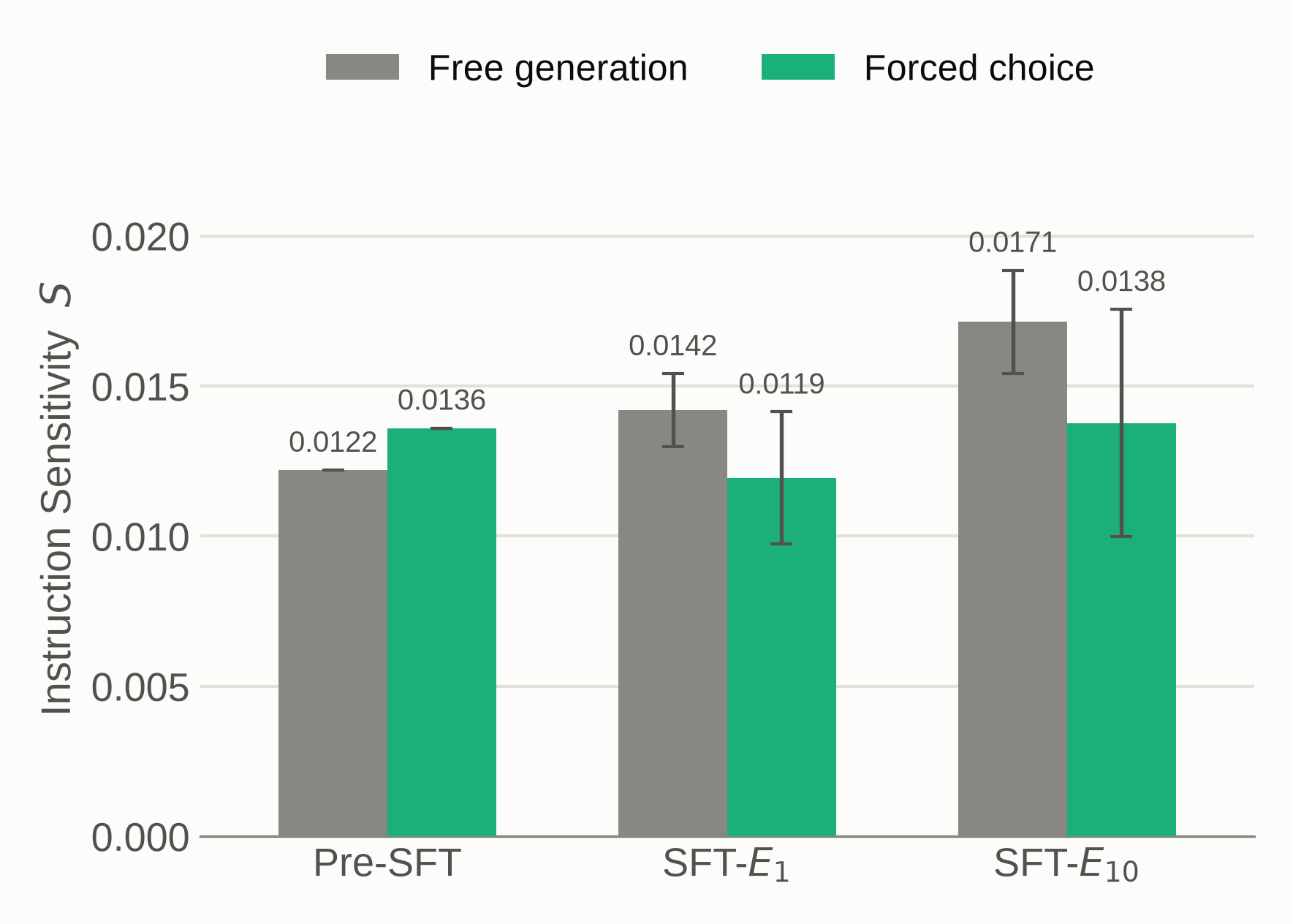}
    \caption{Instruction sensitivity of Qwen3-8B on ESCI-English under free-generation and forced-choice evaluation.
    Free generation shows increased post-SFT sensitivity, whereas the same pattern is not observed under forced-choice likelihood scoring.
    Post-SFT values are averaged over three random seeds.}
    \label{fig:esci_protocol}
\end{figure}

These results indicate that instruction sensitivity can depend on the prediction and scoring procedure itself.
Free generation relies on deterministic autoregressive decoding followed by prefix-based label matching, whereas forced choice directly compares predefined label likelihoods.
Even with nearly perfect label compliance and similar average accuracy, the two procedures yield qualitatively different robustness conclusions.
We therefore interpret this result as evidence of \emph{evaluation-protocol dependence}, rather than as an output-format effect.

This finding also qualifies comparisons with the MS MARCO results.
The apparent post-SFT sensitivity increase on ESCI under free generation should not be interpreted as evidence of the same phenomenon observed in the MS MARCO analysis because it is not reproduced under forced choice.
In contrast, all MS MARCO analyses, including the cross-model comparisons in Section~\ref{sec:results_train_inst}, use the same likelihood-based relevance scoring procedure throughout.
Their scale-, training-instruction-, and cross-model differences are therefore measured under a consistent prediction and scoring protocol.
\section{Discussion}
\label{sec:discussion}

We discuss the relationship between task performance and instruction robustness, the cross-model variation observed in the SFT effect, and the influence of the evaluation protocol on measured instruction sensitivity.

\subsection{Task Performance and Instruction Robustness Are Distinct Outcomes}
\label{sec:discussion_performance_robustness}

The results show that improvements in average task performance do not necessarily imply improved robustness to instruction paraphrases.
Table~\ref{tab:performance_robustness} summarizes mean nDCG@10 and instruction sensitivity for the primary Qwen3 analysis on MS MARCO.
At 1.7B and 4B, the two quantities improve together: SFT increases average ranking performance while substantially reducing sensitivity.
For example, under \(T_A\), mean nDCG@10 increases from \(0.3745\) to \(0.5244\) at 1.7B and from \(0.6383\) to \(0.7240\) at 4B, while sensitivity decreases from \(0.0905\) to \(0.0265\) and from \(0.0332\) to \(0.0143\), respectively.

\begin{table}[t]
\centering
\caption{Mean ranking performance and instruction sensitivity before and after SFT for the primary Qwen3 analysis on MS MARCO. Mean nDCG@10 is computed across the ten evaluation instructions.}
\label{tab:performance_robustness}
\small
\resizebox{\columnwidth}{!}{
\begin{tabular}{llcc}
\hline
Model & Condition & Mean nDCG@10 & Sensitivity \(S\) \\
\hline
1.7B & Pre-SFT      & 0.3745 & 0.0905 \\
     & SFT-\(T_A\) & 0.5244 & 0.0265 \\
     & SFT-\(T_B\) & 0.5015 & 0.0299 \\
\hline
4B   & Pre-SFT      & 0.6383 & 0.0332 \\
     & SFT-\(T_A\) & 0.7240 & 0.0143 \\
     & SFT-\(T_B\) & 0.7096 & 0.0152 \\
\hline
8B   & Pre-SFT      & 0.7096 & 0.0126 \\
     & SFT-\(T_A\) & 0.7405 & 0.0086 \\
     & SFT-\(T_B\) & 0.7293 & 0.0189 \\
\hline
\end{tabular}}
\end{table}

At 8B, however, average effectiveness and robustness no longer move together consistently.
Under \(T_B\), mean nDCG@10 improves from \(0.7096\) to \(0.7293\), while the point estimate of sensitivity increases from \(0.0126\) to \(0.0189\).
Under \(T_A\), mean performance improves to \(0.7405\) while sensitivity decreases to \(0.0086\).
Although neither individual 8B sensitivity change is statistically distinguishable from zero, their paired contrast is statistically reliable.
Thus, successful task adaptation can coexist with different robustness outcomes depending on the training instruction.

The cross-family checks further show that this behavior cannot be explained by model scale alone.
Gemma-2-9B exhibits the same directional contrast as Qwen3-8B, with lower post-SFT sensitivity under \(T_A\) than under \(T_B\) across all three seeds, although the query-bootstrap confidence interval includes zero.
Mistral-7B, in contrast, shows no corresponding training-instruction contrast.
These results suggest that the magnitude and consistency of the robustness effect depend on properties of the adapted model in addition to its parameter scale.

This distinction has an important implication for evaluating task-specific fine-tuning.
Average ranking effectiveness alone may conceal changes in robustness to alternative instruction formulations, while robustness conclusions from one model or training instruction may not transfer directly to another.
Task effectiveness and instruction robustness should therefore be treated as complementary properties when assessing fine-tuned models.

\subsection{Evaluation Protocol Can Change the Apparent Robustness Conclusion}
\label{sec:discussion_protocol}

The ESCI analysis shows that measured instruction sensitivity can depend on how task predictions are obtained and scored.
Under free generation, Qwen3-8B shows increased post-SFT sensitivity, whereas likelihood-based forced choice does not reproduce this pattern.
At the same time, average task performance improves similarly under both protocols.
Thus, the evaluation procedure changes the robustness conclusion without materially changing the conclusion about task effectiveness.

Importantly, this discrepancy cannot be explained simply by failures to follow the required output format.
Across the Pre-SFT and two SFT conditions, the mean valid-label generation rate is \(0.9983\), with only \(0.0027\) standard deviation across the ten evaluation instructions.
Label-format compliance is therefore nearly saturated even under free generation.
Free generation and forced choice should instead be viewed as distinct prediction procedures: the former relies on deterministic autoregressive decoding followed by prefix-based label matching, whereas the latter directly compares the conditional likelihoods of predefined labels.
Our results show that these procedures can yield different estimates of instruction sensitivity even when valid-label generation and average task performance are similar.

This observation is important when instruction sensitivity itself is the quantity of interest.
Robustness estimates obtained under different decoding or scoring procedures should not be assumed to be directly interchangeable.
In contrast, all MS MARCO analyses---including the Qwen3 scale analysis and the Mistral-7B and Gemma-2-9B cross-family checks---use the same likelihood-based relevance scoring procedure across training and evaluation conditions.
The observed scale-, training-instruction-, and cross-model differences are therefore measured under a consistent prediction and scoring protocol.

\subsection{Limitations}
\label{sec:limitations}

This study has several limitations.
First, the systematic scale analysis is limited to three Qwen3 model sizes.
Although Mistral-7B and Gemma-2-9B provide targeted cross-family checks at comparable scales, they do not constitute a full model-family-by-scale analysis.
The Qwen3-8B behavior should therefore not be interpreted as a universal transition point.

Second, the experiments use a limited set of training instructions and ten evaluation paraphrases.
Although the evaluation instructions are designed to preserve task semantics while avoiding exact overlap with the MS MARCO training instructions, broader instruction sets are needed to determine which properties drive different robustness outcomes.
In particular, the three Qwen3-8B conditions do not establish whether \(T_A\) is exceptional or whether the contrasts reflect a broader pattern.

Third, our statistical analysis primarily quantifies query-level uncertainty while treating the trained seeds as observed runs.
For Qwen3-8B, the training-instruction contrasts have consistent directions across all three seeds and remain statistically reliable in every leave-one-seed-out query-bootstrap analysis.
Nevertheless, three seeds are insufficient for strong inference about variability across training runs.

Finally, our results characterize changes in instruction sensitivity but do not identify their underlying mechanism.
The observed differences may reflect task specialization, instruction processing, architectural or pretraining differences, or other effects of fine-tuning.
Further representation- or optimization-level analysis is needed to explain why the robustness effect of SFT varies across models and training instructions.
\section{Conclusion}
\label{sec:conclusion}

We examined how conventional task-specific SFT changes robustness to paraphrased task instructions.
Within the Qwen3 family, instruction sensitivity is lower at larger model scales before fine-tuning.
At 1.7B and 4B, SFT reliably reduces sensitivity under different training instructions.
At 8B, individual sensitivity changes are not statistically distinguishable from zero, but the contrasts between training instructions are statistically reliable.
Cross-family checks provide a more nuanced result: Gemma-2-9B shows the same directional training-instruction contrast as Qwen3-8B, whereas Mistral-7B does not, suggesting that model scale alone does not determine the robustness effect of SFT.
These findings also show that average task performance and instruction robustness need not improve together.

The ESCI analysis further shows that measured sensitivity can depend on the prediction and scoring protocol.
Free-generation and forced-choice evaluation yield similar average performance but qualitatively different sensitivity patterns despite nearly perfect valid-label generation.
This highlights the importance of controlling the evaluation procedure when comparing instruction sensitivity.

Overall, SFT should not be assumed to uniformly reduce instruction sensitivity.
Its robustness effect can vary across model scale, training instruction, and model, while the apparent effect can additionally depend on the evaluation protocol.
Evaluating fine-tuned models across multiple instruction paraphrases and controlled prediction protocols therefore provides a more complete assessment than relying on a single instruction or average performance alone.

\section*{Ethical Considerations}
\label{sec:ethics}

This study analyzes the behavior of pretrained language models under alternative task instructions and does not involve human participants or the collection of personally identifiable information.
All experiments use publicly available benchmark datasets and pretrained open-weight models.
The study does not introduce new user-facing systems or make decisions about individuals.

The main potential concern is that instruction sensitivity may affect the reliability of deployed language-model systems: semantically equivalent instructions can lead to different outputs or performance levels.
Our analysis is intended to improve understanding and evaluation of this behavior rather than to exploit it.
We therefore report robustness across multiple instruction formulations and explicitly examine evaluation artifacts that may otherwise lead to misleading conclusions.

The experiments also require computational resources for repeated fine-tuning and evaluation across model scales and random seeds.
We limit the study to relatively small and medium-sized models and use parameter-efficient fine-tuning to reduce computational cost.
No proprietary or private user data are used.

\bibliographystyle{ACM-Reference-Format}
\bibliography{acmart}











\end{document}